\documentclass[a4paper,twocolumn, secnumarabic, aps, superscriptaddress,
amssymb, amsmath, nobibnotes, showpacs]{revtex4}
\usepackage{graphicx}
\usepackage{dcolumn}

\begin{document}
\title{Facilitated diffusion of DNA-binding proteins: Efficient 
simulation with \\ the method of excess collisions (MEC)}

\author{Holger Merlitz}
 \email{merlitz@gmx.de}
\affiliation{Softmatter Lab, Department of Physics, Xiamen University,
Xiamen 361005, P.R.\ China}
\author{Konstantin V.\ Klenin}
\affiliation{Division of Biophysics of Macromolecules,
German Cancer Research Center, D-69120 Heidelberg, Germany}
\author{Chen-Xu Wu}
\affiliation{Softmatter Lab, Department of Physics, Xiamen University,
Xiamen 361005, P.R.\ China}
\author{J\"org Langowski}
\affiliation{Division of Biophysics of Macromolecules,
German Cancer Research Center, D-69120 Heidelberg, Germany}

\date{\today}

\begin{abstract}
In this paper, a new method to efficiently simulate diffusion 
controlled second order chemical reactions is 
derived and applied to site-specific DNA-binding proteins. The
protein enters a spherical cell and propagates via two
competing modes, a free diffusion and a DNA-sliding mode,
to search for its specific binding site  
in the center of the cell. There is no need for 
a straight forward simulation of this process. Instead, 
an alternative and exact approach is shown to 
be essentially faster than explicit random-walk simulations.
The speed-up of this novel simulation technique is rapidly
growing with system size.
\end{abstract}

\pacs{87.16.Ac}

\maketitle

\section{Introduction} 
Diffusion controlled bio-chemical reactions play a 
central role in keeping any organism alive~\cite{riggs70,richter74}:
The transport of molecules through cell membranes, the passage of
ions across the synaptic gap, or the search carried out by drugs on the way to
their protein receptors are predominantly diffusive
processes. Further more, essentially all of the biological
functions of DNA are performed by proteins that interact
with specific DNA sequences~\cite{berg85, ptashne01},
and these reactions are diffusion-controlled.

However, it has been realized that some proteins
are able to find their specific binding sites on DNA
much more rapidly than is `allowed' by the diffusion
limit~\cite{riggs70, berg81}. It is therefore generally
accepted that some kind of facilitated diffusion must
take place in these cases.
Several mechanisms, differing in details, have been proposed.
All of them essentially involve
two steps: the binding to a random non-specific
DNA site and the diffusion (sliding) along the DNA chain.
These two steps may be reiterated
many times before proteins actually find their target, since the sliding is
occasionally interrupted by dissociation.
Berg~\cite{berg81} and Zhou~\cite{zhou04} 
have provided thorough (but somewhat sophisticated) theories
that allow estimates for the resulting reaction rates.
Recently, Halford has presented
a comprehensive review on this subject and proposed
a remarkably simple and semiquantitative approach
that explicitly contains the mean sliding length
as a parameter of the theory~\cite{halford04}. This 
approach has been refined and put onto a rigorous base
in a recent work by the authors~\cite{klenin05}.

Although analytical models provide a good general
understanding of the problem, they fail to give quantitative
predictions for systems of realistic complexity. Therefore, 
numerical simulations are required to calibrate the set
of parameters that form the backbone of these models.
However, a straight forward simulation of a protein searching
through mega-bases of non-target DNA to find its specific
binding site would be prohibitive for all except for the most 
simple numerical models. Fortunately, there are better ways.
Two of the authors (KK and JL) have recently introduced the method
of excess collisions (MEC) for an efficient simulation of
intramolecular reactions in polymers~\cite{klenin04}. In
the present work, this method is modified to apply to 
second order diffusion controlled chemical reactions 
(Section \ref{sec:mec}). We thereby construct a simple 
random walk approach to facilitated diffusion of DNA-binding
proteins (Section \ref{sec:facilitated}) and apply
the MEC and our analytical estimate for reaction times 
to this model (Section \ref{sec:appl} and \ref{sec:tauf}).
Section \ref{sec:chain} provides details about the 
generation of DNA-chains, followed by a set of
simulations covering a large range of system dimensions
(Section \ref{sec:simu}) to verify the performance of
the MEC.

\section{Theory}  \label{sec:theory}
\subsection{Method of excess collisions (MEC)} \label{sec:mec}
We consider a (time-homogeneous) stochastic process. The problem
is to find the average time $\tau_{\rm BA}$ of the first arrival
at a certain state A, provided that, at time $t=0$, the system
occupied another state B.

Suppose we observe the system for a long time interval $T$ and
monitor the events of entering state A. These events will be
referred to as collisions. Each collision that occurs for the 
first time after visiting state B will be called prime 
collision. We obtain the (asymptotically 
correct for $T\rightarrow \infty$) relation
\begin{equation} \label{eq:5}
T = n(T)\, \tau_R = n'(T)\, \tau'_R\;,
\end{equation}   
where $n(T)$ and $n'(T)$ are the average numbers of all and
of prime collisions during the time interval $T$, respectively,
and $\tau_R$ and $\tau'_R$ are the corresponding mean recurrence times.
Hence,
\begin{equation} \label{eq:10}
\tau'_R = \frac{n(T)}{n'(T)}\, \tau_R \equiv N\,\tau_R\;.
\end{equation}
The ratio $N \equiv n(T)/n'(T)$ defines the average number of
collisions between two visits to state B and does actually
not depend on $T$, once $T$ is chosen sufficiently large. 
The mean recurrence time $\tau'_R$ of prime
collisions is simply the average time the system requires
to move from state A to B and back from state B to A:
\begin{equation} \label{eq:15}
\tau'_R = \tau_{\rm AB} + \tau_{\rm BA}\;,
\end{equation}
where $\tau_{\rm AB}$ is the mean time of first arrival at state
B starting from A. With eq.\ (\ref{eq:10}) we then obtain
\begin{equation} \label{eq:20}
\tau_{\rm BA} = N\tau_R - \tau_{\rm AB}\;.
\end{equation}

This relation is useful for the numerical estimation of $\tau_{\rm BA}$
if $\tau_{\rm BA} \gg \tau_{\rm AB}$. A simulation cycle then starts in state
A and ends as soon as state B is reached, i.e.\ the reversed reaction
$A\rightarrow B$ is simulated in order to obtain the (much lower)
reaction rate of the original reaction $B\rightarrow A$. In this 
case we can write
\begin{equation} \label{eq:25}
N = \langle N_{\rm coll} \rangle + 1\;,
\end{equation}
where $\langle N_{\rm coll} \rangle$ is the average number of collisions
in a simulation cycle and the second term accounts for the prime collision
(which is not observed in the simulations, since the cycle starts at the
time instant that immediately follows the prime collision).
As will be shown later in Section \ref{sec:appl}, the recurrence time
$\tau_R$ can be renormalized and computed efficiently inside
a small test system. Note that eq.\ (\ref{eq:20}) can be written as
\begin{equation} \label{eq:30}
\tau_{\rm BA} \equiv (N_{\rm E} + 1)\, \tau_R\;,
\end{equation}
where
\begin{equation} \label{eq:35}
N_{\rm E} \equiv \langle N_{\rm coll} \rangle - \frac{\tau_{\rm AB}}{\tau_R}
\end{equation}
is the mean number of excess collisions per
simulation cycle~\cite{klenin04}, since the ratio
$\tau_{\rm AB}/\tau_R$
is just the mean number of collisions that would be observed
in a simulation run of length $\tau_{\rm AB}$ with a
starting point at an arbitrary state of the system
(not necessary state A).

\subsection{Simple model for facilitated diffusion
of DNA-binding proteins} \label{sec:facilitated}
We consider a spherical volume (cell) of radius $R$ and
inside it a worm-like chain (DNA) of length $L$ and radius $r_c$.
The protein is represented as a random walker moving inside
the cell with a certain time step $dt$. A collision takes 
place once the walker enters the active binding site, a
spherical volume of radius $r_a$ positioned in the middle of 
the chain that, in its turn,
coincides with the center of the cell.
We want to point out that 
the parameter $r_a$ does not necessary
correspond to any geometrical length in the real system.
It defines a probability for the reaction to take
place, and may cover additional variables which are not
included explicitly in the model, like protein orientation 
and conformation.  An attractive step potential is implemented as
 \begin{equation} \label{eq:40}
U(d) = \left\{
\begin{array}{ccl}
-E_o &\hspace{0.8cm} &d \leq r_c \\
0    &\hspace{0.8cm} &d > r_c \;,\\
\end{array}  
\right. 
\end{equation} 
where $d$ is the shortest distance between walker and chain.
This defines a pipe with radius $r_c$ around the chain contour 
that the walker is allowed to enter freely from outside, but to exit 
only with the probability    
\begin{equation} \label{eq:45}
p = \exp (-E_o/k_{\rm B} T) \;,
\end{equation}
where $k_{\rm B} T$ is the Boltzmann factor, otherwise it is 
reflected back inside the chain. We
may therefore denote $p$ as {\it exit probability}.
It is important to note that $p$ defines the equilibrium
constant $K$ of the two phases, the free and the non-specifically
bound protein, according to 
\begin{equation} \label{eq:50}
K \equiv  \frac{\sigma}{c} = \frac{\pi\, r_c^2}{p}\;,
\end{equation}
where $c$ is the concentration of free proteins and
$\sigma = c\,V_c/(p\,L)$ is the linear density of proteins that are
non-specifically
bound to the DNA, with
$V_c = \pi\, r_c^2\, L$ being the geometric volume of the chain.

\subsection{Method of computation of the recurrence time}
\label{sec:appl}
The two states of interest are the protein entering the cell,
B, and the same protein reaching the active site in the center
of the cell, A. More specifically, we are interested in finding the
time $\tau_{\rm BA}$ the walker requires to reach a distance
$r = r_a$ when starting at distance $r(t=0) = R$.

We shall first define the excluded volume of the chain as
\begin{equation} \label{eq:55}
V_{\rm ex} \equiv \int_{V}
\left[1 - \exp\left(\frac{-U[d({\bf r})]} {k_{\rm B} T}\right)\right]\, d{\bf
r}
= V_c\, \left( 1 - \frac{1}{p} \right)\;,
\end{equation}
where $U(d)$ is the energy of the walker
as defined by eq.\ (\ref{eq:40}) and the integration is performed
over the geometric volume of the cell, $V = (4/3) \pi R^3$.
The effective volume $V_{\rm eff}$ of the cell is then
\begin{equation} \label{eq:65}
V_{\rm eff} \equiv V - V_{\rm ex} = V + V_c\, \left(\frac{1}{p} - 1 \right)\;.
\end{equation}
Next we assume that simulations were carried out within
a small test system of radius $R^* < R$ and that 
the recurrence time $\tau^*_R$ of the walker was found. 
Its recurrence time in the larger system is then found as
\begin{equation} \label{eq:80}
\tau_R(V) = \tilde{\tau}_R\, V_{\rm eff}\;,
\end{equation} 
where we have defined
\begin{equation} \label{eq:85}
\tilde{\tau}_R \equiv \frac{\tau^*_R}{V^*_{\rm eff}}\;.
\end{equation} 
This ratio does not depend on system size and 
may therefore be called 
{\it specific recurrence time}. It only depends on the
potential-depth $E_o$ and the step-size chosen for the 
random walk. The idea is to compute $\tilde{\tau}_R$ 
(as described in Section \ref{sec:tautilde}) 
for a small test system with dimensions of the order of $r_a$
(which is the radius of the
specific binding site) to obtain $\tau_R$ for the system of interest
using eq.\ (\ref{eq:80}).
Once $\tau_R$ is known, $\tau_{\rm AB}$ is computed via random walk 
simulations in the large system, starting at $r(t=0) = r_a$
and terminating as soon as the periphery of the cell $r(\tau_{\rm AB}) = R$
is reached. Following the trajectory of the walker, the number of 
collisions $\langle N_{\rm coll} \rangle = N - 1$ is
monitored as well, so that eq.\ (\ref{eq:20}) can be used
to determine the much longer reaction time $\tau_{\rm BA}$.

\subsection{Analytical estimate for the collision time}
\label{sec:tauf}
As has been discussed in detail elsewhere~\cite{klenin05},
it is possible to estimate the reaction time for the protein
using an analytical approach, once certain
conditions are satisfied. The resulting expression is
\begin{equation} \label{eq:95}
\tau_{\rm BA}(\xi) = \left( \frac{V}{8D_{\rm 3d}\,\xi} +
    \frac{\pi\,L\, \xi}{4 D_{\rm 1d}} \right) \left[
    1 - \frac{2}{\pi} \arctan \left(\frac{r_a}{\xi}\right)\right]\;
\end{equation}
with the 'sliding' variable
\begin{equation} \label{eq:100}
\xi = \sqrt{\frac{D_{\rm 1d}\, K}{2\pi\, D_{\rm 3d}}}\;
\end{equation}
and $D_{\rm 1d}$ and $D_{\rm 3d}$ being the diffusion coefficients
in sliding-mode and free diffusion, respectively. Generally,
the equilibrium constant $K$ has to be determined in simulations
of a (small) test system, containing a piece of chain without specific
binding site~\cite{klenin05}. In the present model, $K$ is
known analytically via eq.\ (\ref{eq:50}). If the 
step-size $dr$ of the random walker is 
equal both inside and outside the chain
(the direction of the step being arbitrary), we further have
$D_{\rm 1d} = D_{\rm 3d} = dr^2/6$, and hence obtain
\begin{equation} \label{eq:105}
\xi = \sqrt{\frac{r_c^2}{2p}}\;.
\end{equation}
This variable has got the dimension of length; as we have pointed
out in~\cite{klenin05}, it corresponds to the average sliding 
length of the protein along the DNA contour in Halford's 
model~\cite{halford04}. In this light, a (non rigorous) 
interpretation of eq.\ (\ref{eq:95}) is as follows: 
The first term in the round brackets represents
the time of free diffusion of the walker,
whereas the second term stands for the time of one-dimensional sliding.
With increasing affinity of the walker to the chain (expressed
as a reduced value for the exit probability $p$), the 
sliding variable $\xi$ increases and the contribution
of free diffusion to the reaction time (first term in \ref{eq:95})
becomes less significant. At the same time, the second term
of eq.\ (\ref{eq:95}) is growing. Depending on the choice of
system parameters, there may be a turning point where the
latter contribution over-compensates the former, so that the 
total reaction time increases once $\xi$ is growing further.

For a random walk model as simple as used here,
this analytical formula describes the reaction times well
within 10\% tolerance, as long as the
following conditions are satisfied: (1) $\xi \ll R$, i.e.\ the
sliding parameter should be small compared to the system size.
This restriction assures the correct normalization of the
protein's probability distributions and the diffusion 
efficiencies as discussed in~\cite{klenin05}. 
(2) During the diffusion process, the system reaches its 
equilibrium, so that the
constant $K$ represents the average times the protein spends
in free and in non-specifically bound mode. This requires either a
crowded environment (the chain-density inside the cell is high
enough) or a reasonably small value for $\xi$, since the
initial position of the walker is always at the periphery
and outside the chain, i.e.\ not in equilibrium. (3) $\xi < l_p$,
where $l_p$ is the persistence length of the chain. This 
restriction accounts for the assumption that the walker moves
along an approximately straight line during one sliding period. 
However, numerical tests have shown that deviations from a straight
geometry actually have little impact to the accuracy of the model.
(4) The step-size of the random walk has to be small compared to
the size of the binding site. 

It should be pointed out that an analytical approach as simple
as that is by no means 
supposed to simulate the actual situation in a living cell.
Instead, it serves as a platform for a much wider
class of semi-empirical models. The sliding-parameter $\xi$ 
contains the affinity of non-specific
protein-DNA binding and is flexible to vary with the potential
chosen for the simulation. 
The diffusion coefficients $D_{\rm 1d}$
and $D_{\rm 3d}$ can be adapted to experimental measurements,
and the target size $r_a$ contains protein-specific reaction
probabilities. These parameters can be fitted to
either describe system-specific experimental results or
the output of more sophisticated numerical codes which
would otherwise not permit any analytical treatment.

\section{Numerical Model}   \label{sec:chain}
In order to approximate the real biological situation,
the DNA was modeled by a chain of straight segments of equal length $l_0$.
Its mechanical stiffness was defined by the bending energy
associated with each chain joint:
\begin{equation} \label{eq:110}
E_b = k_{\rm B} T \, \alpha \, \theta^2\;,
\end{equation}
where $\alpha$ represents the dimensionless
stiffness parameter, and $\theta$ the bending angle. The numerical
value of $\alpha$ defines the persistence length ($l_p$), 
i.e.\ the ``stiffness'' of the chain.
The excluded volume effect was taken into account by introducing
the effective chain radius $r_c$. The conformations of the chain,
with distances between non-adjacent segments smaller than $r_c$,
were forbidden.
The target of specific binding was assumed to lie exactly in the
middle of the DNA.
The whole chain was packed in a spherical volume (cell) of radius $R$
in such a way that the target occupied the central position.

To achieve a close packing of the chain inside the
cell, we used the following algorithm. First,
a relaxed conformation of the free chain was produced
by the standard Metropolis Monte-Carlo (MC) method.
For the further compression, we defined the
center-norm (c-norm) as the maximum distance from the target
(the middle point) to the other parts of the chain.
Then, the MC procedure was continued with one modification.
Namely, a MC step was rejected if the
c-norm was exceeding 105\% of the lowest value registered so
far. The procedure was stopped when the desired degree of compaction
was obtained.

The protein was modeled as a random walker within the cell
with reflecting boundaries.
During one time-step it was displaced by the
distance $dr$ in a random direction.
Once approaching the chain closer than its radius
$r_c$ defining the ``non-specific binding pipe", 
it was allowed to enter it freely and continue
its random walk inside. Upon crossing the pipe boundary 
from inside, it was either allowed to pass with the 
exit probability $p$ or otherwise reflected
back inside, as described in Section \ref{sec:facilitated}. 

Below in this paper, one step $dt$ was chosen as the unit of time and
one persistence length $l_p = 50$ nm of the DNA chain as the unit of distance.
The following values of parameters were used. The length of one segment
was chosen as $l_0 = 0.2$, so that one persistence length was partitioned into
5 segments. The corresponding value of the stiffness parameter was
$\alpha = 2.403$~\cite{klenin98}.
The chain radius was $r_c = 0.06$, and the active site
was modeled as a sphere of identical radius $r_a = 0.06$ embedded 
into the chain. The step-size of the random walker both
inside and outside the chain was $dr = 0.02$, corresponding to a diffusion 
coefficient $D_{\rm 3d} = D_{\rm 1d} = dr^2/6 = 2 \cdot 10^{-4}/3$.
This choice was a compromise between accuracy and simulation 
time. Tests have confirmed that a smaller step-size could somewhat
reduce the gap between theoretical (eq.\ \ref{eq:95}) and
simulated reaction time at small values of $\xi$.

\section{Computation of the specific recurrence time}
\label{sec:tautilde}
To compute the specific recurrence time $\tilde{\tau}_R$ of
eq.\ (\ref{eq:85}), a very small test system is sufficient.
Moreover, the computations can be carried out for the collisions
from within the specific binding site of radius $r_a$~\cite{klenin04}.
The entire
system, i.e.\ the sphere and a short piece of chain, was
embedded into a cube of $4 r_a$ side-length with reflective 
walls. In principle, the size of the cube should be of no 
relevance, but it was found that, if chosen too small,
effects of the finite step-size were emerging. The walker
started inside the sphere. Each time upon leaving the spherical
volume a collision was noted. If the walker was about to exit
the cylindrical volume of the chain, it was reflected
back inside with the probability $1-p$.
The clock was halted as long as the walker moved outside
the sphere and only counted time-steps inside the sphere.
Since the binding site was embedded into the chain, its
effective volume (eq.\ \ref{eq:65}) was simply  
$V_{\rm eff} = V_a/p$, with $V_a$ being the volume of the 
specific binding site.

\begin{table}[b]
\caption{Recurrence time (3rd column) inside the spherical binding
site ($R = r_a$), specific recurrence time eq.\ (\protect\ref{eq:85})
(4th column), and simulation results for the large system ($R = 4.8$,
column 5-9).
The first column is the exponent of the exit probability $p = 2^{-l}$,
the second column the corresponding sliding parameter eq.\ 
(\protect\ref{eq:105}). The last
column defines the speed-up achieved with the MEC approach.
\label{tab:ncoll}}
\begin{center}
\begin{tabular}{r|c|cc|ccccc}
\multicolumn{2}{c|}{} &  \multicolumn{2}{c|}{$R = r_a$} & 
\multicolumn{5}{c}{ $R = 4.8$ } \\ \hline
$l$ & $\xi $&$\tau^*_R$ & $\tilde{\tau}_R$ & $N$ & $\tau_{\rm AB}$& 
$\tau_{\rm BA}$(MEC) & $\tau_{\rm BA}$ & $\frac{\tau_{\rm BA}}{\tau_{\rm AB}}$
\\ \hline
0 & 0.042 & 4.039 & 4464 & 4.928 & 58577 & $1.013\cdot 10^7$ & $1.029\cdot 10^7$ & 176  \\
1 & 0.060 & 4.693 & 2594 & 7.019 & 58674 & $8.445\cdot 10^6$ & $8.131\cdot 10^6$ & 139  \\
2 & 0.085 & 5.112 & 1413 & 10.88 & 59484 & $7.243\cdot 10^6$ & $6.818\cdot 10^6$ & 115  \\
3 & 0.120 & 5.368 & 741.6& 16.05 & 61225 & $5.776\cdot 10^6$ & $5.823\cdot 10^6$ & 95.1 \\
4 & 0.170 & 5.496 & 379.7& 25.66 & 65418 & $5.020\cdot 10^6$ & $4.876\cdot 10^6$ & 74.5 \\
5 & 0.240 & 5.575 & 192.6& 39.50 & 75501 & $4.370\cdot 10^6$ & $4.272\cdot 10^6$ & 56.6 \\
6 & 0.339 & 5.606 & 96.81& 58.56 & 90422 & $3.933\cdot 10^6$ & $3.982\cdot 10^6$ & 44.0 \\
7 & 0.480 & 5.631 & 48.62& 86.29 & 115401& $3.911\cdot 10^6$ & $3.815\cdot 10^6$ & 33.1 \\
8 & 0.679 & 5.629 & 24.30& 122.8 & 172755& $4.184\cdot 10^6$ & $4.119\cdot 10^6$ & 23.8 \\
9 & 0.960 & 5.638 & 12.17& 179.7 & 273757& $5.110\cdot 10^6$ & $5.018\cdot 10^6$ & 18.3 \\
10& 1.358 & 5.642 & 6.089& 253.1 & 422792& $6.456\cdot 10^6$ & $6.243\cdot 10^6$ & 14.8 \\
11& 1.920 & 5.640 & 3.044& 357.1 & 701443& $8.502\cdot 10^6$ & $8.616\cdot 10^6$ & 12.3 \\

\end{tabular}
\end{center}
\end{table}

Table \ref{tab:ncoll} contains the results for 12 
different values of the exit probability $p$. The recurrence
time $\tau^*_R$ does in fact depend on $p$, although
the spherical volume $V_a$ is fully embedded into the chain.
The reason is that within one time-step, the walker may 
leave the sphere, but, depending on $p$, subsequently 
reflected back from the chain's periphery into the
binding site. Such a move is not accounted as a
collision (there are no fractional time-steps). 
The computational cost of these simulations is
negligible --- Millions of cycles are carried out within 
minutes on a PC, and the statistical error of $\tilde{\tau}_R$
can be made negligibly small.

\begin{figure}[t]
\centerline{
\includegraphics[width=1.0\columnwidth]{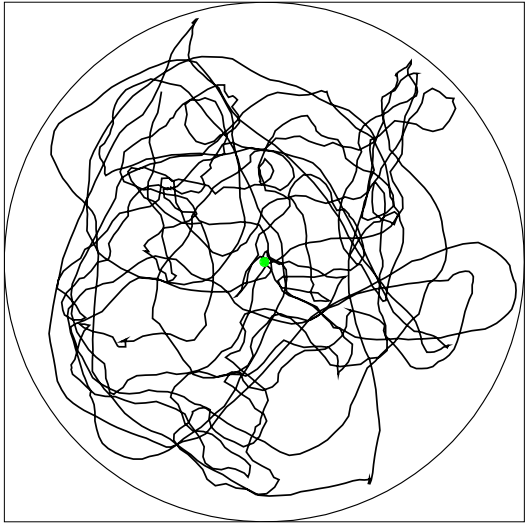}
}
\caption{A `cell' of radius $R=4.8$  (persistence lengths)
containing a chain of $L = 345.8$, corresponding to 240 nm and
17.3 $\mu$m, respectively. The chain was made of 1729 segments.
The protein's specific binding site
is located at the center (dot, not to scale). 
\label{fig:chain120}}
\end{figure}

\section{Model systems of various sizes}
\label{sec:simu}
Next, simulations were carried out for cells of different 
volumes $V_i = 4\pi\,R_i^3/3$
(see table \ref{tab:simu} for 
a summary of the system parameters). The chain lengths
$L_i$ were chosen so that the density $L_i/V_i$ remained of the same order
around 3/4. First, the chain conformation was generated using
the procedure of Section \ref{sec:chain}. Then,
each simulation cycle
started at the periphery of the active binding site (state A)
and ended as soon as the periphery of the cell (state B) was reached.
Whenever the walker returned back to the binding site 
($r < r_a$), one collision
was noted. As long as the walker remained inside the binding
site, the clock was halted. For each value of the exit parameter
$p$, which is related to the walker-chain affinity via
eq.\ (\ref{eq:45}), 2000 cycles
were carried out and the measurements were averaged, so that
statistical fluctuations were reduced to about
2\%. The simulations provided measurements of $\tau_{\rm AB}$, the
average time to reach B when starting from A, and 
$\langle N_{\rm coll} \rangle$, the number of returns to
A on the way towards B. Equations (\ref{eq:80}) and (\ref{eq:20}),
which form the core of the 
MEC approach, were then applied to evaluate $\tau_{\rm BA}$.
Additionally, $\tau_{\rm BA}$ was simulated explicitly,
starting from B, as a verification of the
speed-up and accuracy of the MEC approach. The results are
summarized in table \ref{tab:simu}. In order to clarify
the procedure, we shall first discuss the simulation
of the largest cell $R = 4.8$ in more detail. 

\begin{figure}[t]
\centerline{
\includegraphics[width=1.0\columnwidth]{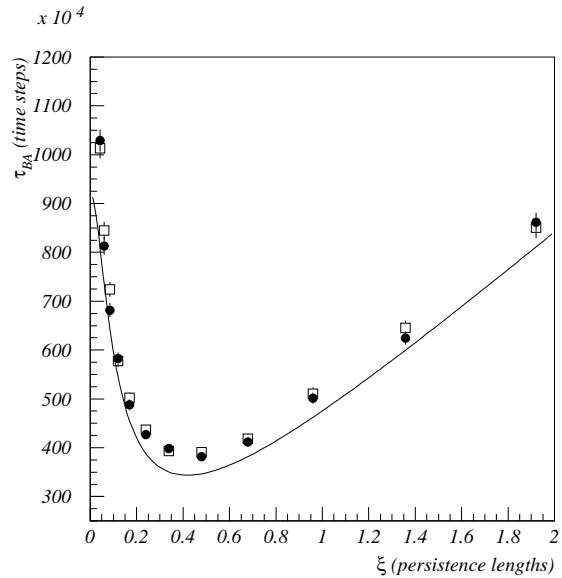}
}
\caption{The first reaction time $\tau_{\rm BA}$ for the cell
of radius $R = 4.8$ persistence lengths as a function of
the sliding parameter $\xi$. Dots: explicit
simulation. Squares: MEC approach, which is exact
within statistical errors. Speed-up: Factor $33.4$ of 
simulation steps after integration over all data points. 
The solid curve is
the analytical estimate eq.\ (\protect\ref{eq:95}).
\label{fig:tauf}}
\end{figure}

Figure \ref{fig:chain120} displays the chain conformation inside
the spherical cell in a 2-dimensional projection. The specific
binding site is located at the center of the cell. Note that,
wherever possible, the chain contour, constructed of 1729
cylindrical segments, tries to avoid large bond angles,
a result of the bending potential as discussed in Section
\ref{sec:chain}.

Table \ref{tab:ncoll} contains details of the simulation results
for 12 different values of the exit parameter $p$,
varied as $p = 2^{-l}$, $l = 0, \dots, 11$. The second
column is the sliding parameter eq.\ (\ref{eq:105}). With
increasing protein-chain affinity, the
walker is spending more time inside the chain volume
so that the sliding parameter is growing in size, reaching
a value of almost two persistence lengths at $p = 2^{-11}$.
The following two columns are the recurrence time $\tau^*_R$
and $\tilde{\tau}_R$ as discussed in Sec.\ \ref{sec:tautilde}.
The next column is the number of collisions $N$ 
(eq.\ \ref{eq:25}). The more time it spends inside the
chain contour, i.e.\ with increasing influence of facilitated
diffusion, the more often the walker returns back to state A
to cause a collision, before being able to reach state B for the
first time to finish the cycle. From $p = 1$ (free diffusion) to
$p = 2^{-11}$, the value of $N$ gains almost two orders 
of magnitude. The next column is the average reaction
time $\tau_{\rm AB}$ of the direction $A \rightarrow B$. This
quantity initially remains almost constant,
but at higher values of protein-chain affinity it 
begins to grow rapidly. The reason is because the walker 
becomes more and more trapped inside the chain volume
and is unable to access the cell periphery as effectively
as it does during free diffusion. The next column is
the reaction time $\tau_{\rm BA}$ of the reaction 
$B \rightarrow A$ as delivered by the
MEC approach using eq.\ (\ref{eq:20}). The recurrence
time $\tau_R$ was determined using eq.\ (\ref{eq:80}),
with the effective volume of eq.\ (\ref{eq:65}) and
the specific recurrence time $\tilde{\tau}_R$ 
(column 4). The next column
contains $\tau_{\rm BA}$ as obtained by direct simulations.
When averaged over all data points, both results for 
$\tau_{\rm BA}$ differed by 2.4\%.
As shown in the last column, the ratio $\tau_{\rm BA}/\tau_{\rm AB}$
was of the order 10-100.
This defines the speed-up of the MEC approach
over the explicit simulation of $\tau_{\rm BA}$. Integrated
over all data points, the total speed-up was equal to $33.4$.

\begin{table}[b]
\caption{Simulation parameters (cell radius $R$, chain length $L$)
and total speed-up. 
Column 3 contains the total number of time-steps n(BA) 
(integrated over all data points)
for the explicit simulation of $\tau_{\rm BA}$, column 4 is the 
integrated speed-up of MEC (the ratio n(BA)/n(AB)).
The last column contains the deviation (averaged over all data
points) between $\tau_{\rm BA}$(explicit) and $\tau_{\rm BA}$(MEC).
\label{tab:simu}}
\begin{center}
\begin{tabular}{c|cccc}
Cell $R$     & Chain $L$  & Time-steps & Speed-up &  Error (\%)  \\ \hline
  1.2   &   5.40    &    $1.68\cdot 10^9$     &   2.3  & 3.9    \\
  2.0     &  25.0     &    $9.31\cdot 10^9$     &   6.9  & 3.9  \\
3.2     &   102.6  &    $4.17\cdot 10^{10}$  &   16.6 & 2.3    \\
4.8     &   345.8  &    $1.44\cdot 10^{11}$  &   33.4 & 2.4     \\
\end{tabular}
\end{center}
\end{table}

Figure \ref{fig:tauf} displays the first reaction times
$\tau_{\rm BA}$ as a function of the sliding parameter $\xi$.
Both methods (explicit simulation and MEC approach) 
deliver identical results within the statistical errors.
The solid curve is a plot of the analytical estimate
eq.\ (\ref{eq:95}), which consistently under-estimates 
the first reaction time by 5-10\% but otherwise
describes the trends accurately, including the location
of the minimum. The results prove
that facilitated diffusion is able to accelerate
the reaction considerably. It is also obvious that
a very high affinity of the protein to the chain 
becomes counter-productive: The walker spends long 
periods of time trapped within a particular loop
of the chain without being able to explore the 
remaining parts of the cell exhaustively. Ideally, the affinity
has to be chosen so that the walker is occasionally able 
to dissociate from the chain and associate again after having
passed some time in free diffusion. The actual
value of the ideal affinity depends on the system
parameters and is easily estimated using eq.\ (\ref{eq:95})
prior to any simulations.      
  
Table \ref{tab:simu} contains a summary of the simulation
results for various system sizes. It appears that
the speed-up delivered by the MEC approach increased
proportional to the square of the cell radius, and 
gained a significant dimension in the largest of our 
test systems. Whereas a cell as small as $R = 1.2$
was treated within 30 minutes on a PC, including
2000 runs of explicit simulation $B\rightarrow A$
for 12 different values of the exit probability $p$, 
the large
cell of $R = 4.8$ required more than 5 days for the same
set of computations. The MEC method reduced 
that time to less than four hours.

\section{Summary}
In this work, the method of excess-collisions (MEC),
recently introduced as a technique to speed up
the simulation of intramolecular reactions in
polymers, is generalized to second order diffusion
controlled reactions, and applied to the problem
of facilitated diffusion of site-specific DNA-binding 
proteins. This method is based on
eq.\ (\ref{eq:20}) and (\ref{eq:80}) to simulate
the much faster back-reaction $A \rightarrow B$ (protein
starts at the binding site and propagates to the
cell-periphery) instead of $B \rightarrow A$.   
We have demonstrated how MEC led to a speed-up 
of up to two orders of magnitude, depending on 
protein-DNA affinity (Table \ref{tab:ncoll}),
and gaining significance with increasing cell 
size (Table \ref{tab:simu}).

The cell model employed in this work was perhaps
the most simple ansatz that was possible without 
being trivial, and 
intentionally so. The simulations had to cover a 
large range of system sizes in order to verify the 
efficiency of the MEC approach. The
chain-lengths span a factor of 64 from the smallest to
the largest system. Nevertheless, the validity of our
results does not depend on the complexity of the model,
such as protein-DNA potential, which modifies the 
equilibrium constant $K$ in eq.\ (\ref{eq:50}) and
thereby the sliding parameter $\xi$ (eq.\ \ref{eq:100}),
hydrodynamic interactions, which would lead to
effective diffusion coefficients, also modifying
$\xi$, or the introduction of protein orientation
and conformation, acting on the effective target 
size $r_a$. The speed-up is consistently evaluated
in terms of simulation steps, not CPU-time, to
ensure invariance on the complexity of the underlying 
protein/DNA model. Based on the results presented here,
the MEC approach can be expected to reduce the numerical effort 
by orders of magnitude,
once more sophisticated (and time consuming) simulation
techniques are employed to study biochemical reaction times
in systems of realistic dimensions.

\end{document}